\documentclass[letter,
               keeplastbox,   % flushend option: not to un-indent last line in References
               ]{jacow}
\usepackage{pdfpages,multirow,ragged2e} %
\makeatletter%
	\ifboolexpr{bool{xetex}}
	 {\renewcommand{\Gin@extensions}{.pdf,%
	                    .png,.jpg,.bmp,.pict,.tif,.psd,.mac,.sga,.tga,.gif,%
	                    .eps,.ps,%
	                    }}{}
\makeatother

\ifboolexpr{bool{xetex} or bool{luatex}} % test for XeTeX/LuaTeX
 {}                                      % input encoding is utf8 by default
 {\usepackage[utf8]{inputenc}}           % switch to utf8

\usepackage[USenglish]{babel}

\ifboolexpr{bool{jacowbiblatex}}%
 {%
  \addbibresource{jacow-test.bib}
  \addbibresource{biblatex-examples.bib}
 }{}
\begin{document}

\title{Fermilab Booster beam emittances from\\ quadrupole modes measured
  by \NoCaseChange{BPMs}\thanks{This manuscript has been authored by
    Fermi Research Alliance, LLC under Contract No. DE-AC02-07CH11359
    with the U.S. Department of Energy, Office of Science, Office of
    High Energy Physics.}}

\author{C. Y. Tan\thanks{cytan@fnal.gov}, M. Balcewicz, Fermi National
  Accelerator Laboratory, Batavia, IL60510, USA 
}
	
\maketitle

\begin{abstract}
   The measurement of beam emittances by extracting the quadrupole mode
  signal from a 4 plate beam position monitor (BPM) was published at least 40 years
  ago. Unfortunately, in practice, this method suffers from poor
  signal to noise ratio and requires a lot of tuning to extract out
  the emittances. In this paper, an improved method where multiple
  BPMs are used together with better mathematical analysis is
  described. The BPM derived emittances are then compared with those
  measured by the Ion Profile Monitor (IPM). Surprisingly, the BPM
  measured emittances behave very well and are more realistic than
  those measured by the IPM.
\end{abstract}

\section{Introduction}
The measurement of beam emittances by extracting the quadrupole mode
signal from a 4 plate beam position monitors (BPM) was published at
least 40 years ago \cite{miller, nassibian}. However, the quadrupole
signal is very small when compared to the dipole signal and so, in
practice, this method suffers from poor signal to noise \cite{russel}. This means
that to actually get the emittance, a lot of tuning is required. We
decided a revisit of this method because the Booster Ionization Profile
Monitors (IPMs) have idiosyncrasies that are very puzzling
\cite{bhat}. For example, the beam current measured by the IPM is not
conserved. It also shows an emittance growth during the
Booster ramp that is unexplained \cite{shiltsev}.

Our contribution to the improvement of this method are the use of
multiple BPMs which then gives us a better way for emittance
extraction. When we compared the emittances extracted with our method
to those measured by the IPMs in the Booster, we found that the BPM measured emittances were more realistic than
those from the IPMs.

\section{Theory}

This section only contains an abridged version of the theory. A full
derivation can be found in Ref.~\cite{cytan} which follows Miller \cite{miller}.

\begin{figure}[!htb]
   \centering
   \includegraphics*[width=.7\columnwidth]{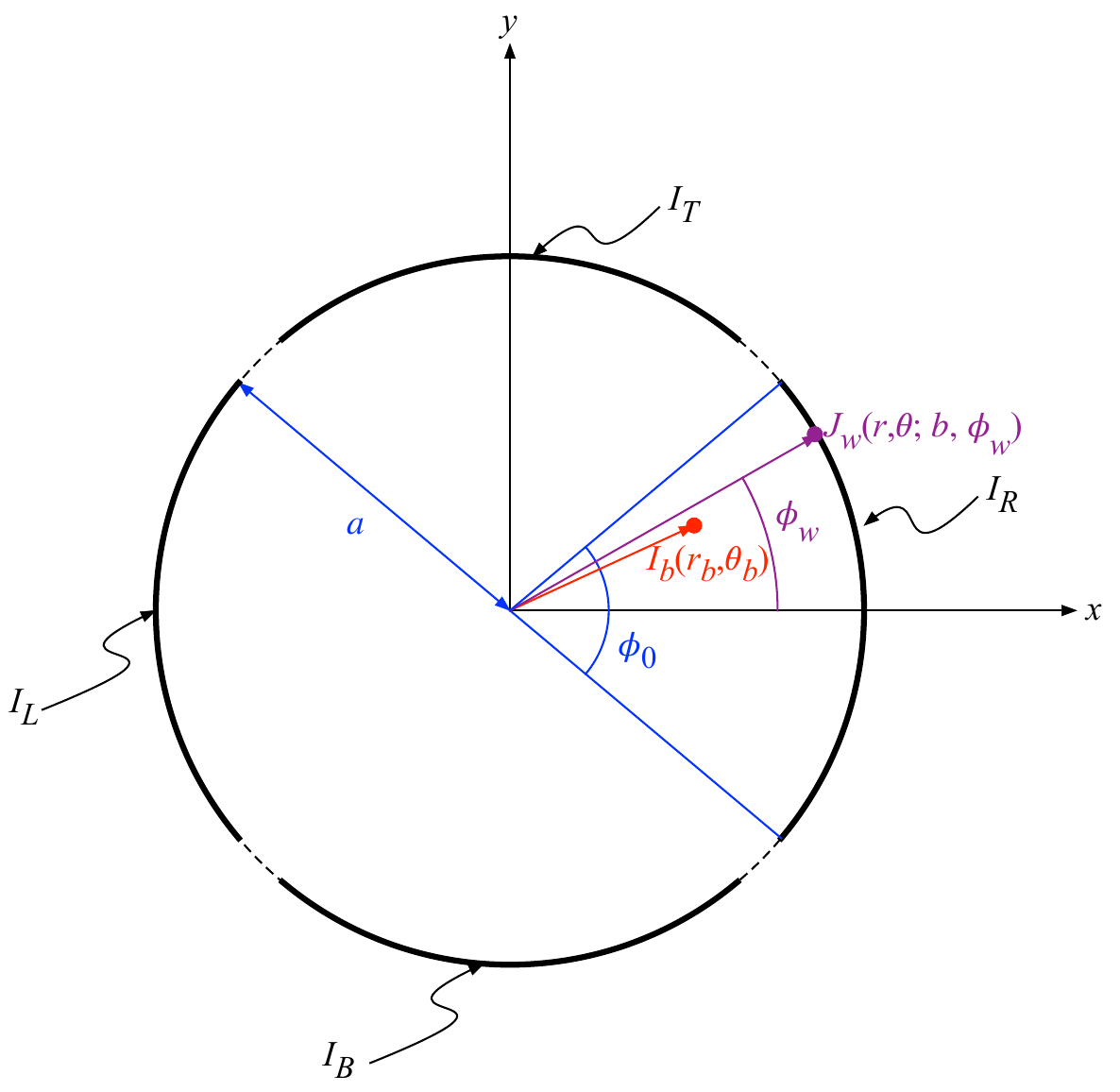}
   \caption{This cartoon shows the line current $I_b$ at
     $(r_b, \theta_b)$ inducing a current density on the $R$(ight) pickup
     plate. All the plates have radius $a$ and subtends an angle
     $\phi_0$. The current density at $(a,\phi_w)$ is $J_w$. }
   \label{setup.pdf}
\end{figure}

The cross section of the 4 plate BPM is shown in
Fig.~\ref{setup.pdf}. The image current density, $J_w$, that is
induced by a pencil current, $I_b$, at $(r_b,\theta_b)$ is given
by
\begin{equation}
  \begin{aligned}
    & J_w(r_b,\theta_b; b, \phi_w) = \\
    &\quad-\frac{I_b(r_b,\theta_b)}{2\pi a}
    \left[
1 + 2\sum_{n=1}^\infty\left(\frac{r_b}{a}\right)^n    \cos n\left(\phi_w -\theta_b\right)
\right].
\end{aligned}
\label{th.1}  
\end{equation}
We can integrate the above to obtain the current on each plate $R$(ight), 
$L$(eft), $T$(op) and $B$(ottom). For example the current on the $R$(ight) plate, $I_R$, is
simply $I_R = \int_{-\phi_0/2}^{\phi_0/2}d\phi\; a J_w$ to give
\begin{equation}
  \begin{aligned}
  I_R &= 
  -\frac{I_b(x_b,y_b)}{2\pi}\left\{
  \phi_0 + 2\left[2\left(\frac{x_b}{a}\right)   \sin\frac{\phi_0}{2}
    +
\right.\right.
\\
&\quad
\left.\left.
  \left(\frac{x^2_b-y^2_b}{a^2}\right) \sin\phi_0
    \right]
  \right\}.
\end{aligned}
\label{th.2}
\end{equation}
where we have only kept terms lower than $(r_b/a)^3$.  
We can obtain similar equations for $I_L$, $I_T$ and $I_B$.

And if the transverse distribution of the beam is a bi-gaussian
distribution centred at $(\bar x, \bar y)$ with standard deviations
$\sigma_x$ and $\sigma_y$ in the $x$ and $y$ directions respectively,
then the normalized gaussian distribution function is
\begin{equation}
    \rho(x,y) = \frac{1}{2\pi\sigma_x\sigma_y}\exp\left[-\frac{(x-\bar x)^2}{2\sigma_x^2}\right]
    \exp\left[-\frac{(y-\bar y)^2}{2\sigma_y^2}\right].
    \label{th.3}
  \end{equation}
With the above density function, the new current distribution on the
$R$(ight) plate
is
\begin{equation}
  \begin{aligned}
    R =  \int_{-\infty}^\infty dx\; dy\; \rho I_R,&=-\frac{I_b}{2\pi}
    \left\{
    \phi_0 + 2\left[
      2\left(\frac{\bar x}{a}\right)\sin\frac{\phi_0}{2}
      +
    \right.\right.\\
  &
  \quad
  \left.\left.
        \left(\frac{\sigma_x^2-\sigma_y^2}{a^2} +\frac{\bar x^2-\bar y^2}{a^2}\right)\sin\phi_0
      \right]
      \right\}.
    \end{aligned}
    \label{th.4}
  \end{equation}
  The remaining plates will also have a similar current distributions
  $L$, $T$, $B$.

  Finally, we can take the appropriate sum and difference combinations
  of $R$, $L$, $T$ and $B$ to create the dipole modes, $d_x$ and $d_y$
  and quadrupole mode $q$
  \begin{equation}
    \left.
    \begin{aligned}
    d_x &= \frac{R-L}{R+L+T+B} =
    \frac{\sin\frac{\phi_0}{2}}{\frac{\phi_0}{2}}\left(\frac{\bar
        x}{a}\right)\\
d_y &= \frac{T-B}{R+L+T+B} =
    \frac{\sin\frac{\phi_0}{2}}{\frac{\phi_0}{2}}\left(\frac{\bar
        y}{a}\right)\\
    q & =\frac{2\sin\phi_0}{\phi_0}\left(
      \frac{\sigma_x^2-\sigma_y^2}{a^2} + \frac{\bar x^2-\bar y^2}{a^2} 
    \right)
  \end{aligned}
  \quad\right\}. 
  \label{th.5}
\end{equation}
$q$ can be written in terms of $d_x$ and $d_y$ to give
\begin{equation}
      q - ( d_x^2 -  d_y^2)\phi_0\cot\frac{\phi_0}{2}=
      \frac{2\sin\phi_0}{\phi_0}\left(\frac{\sigma_x^2-
          \sigma_y^2}{a^2}\right).
      \label{th.6}
\end{equation}  
The above can be written In terms of lattice functions by using the
following relationship
\begin{equation}
  \sigma_{x,y}^2 = \frac{\beta_{x,y}\epsilon_{x,y}}{\pi} +
  (D_{x,y}\sigma_p)^2
  \label{th.7}
\end{equation}  
where $\epsilon_{x,y}$ are the emittances, $\beta_{x,y}$ are the beta
functions, $D_{x,y}$ are the dispersions of the beam in the $x$ and
$y$ directions respectively; and
$\sigma^2_p =\langle (dp/p)^2\rangle $ is the standard deviation of
the relative momentum spread of the beam.

Thus, Eq.~(\ref{th.6}) becomes
\begin{equation}
  \Delta_q
\equiv \frac{2}{a^2}\frac{\sin\phi_0}{\phi_0} \left[
    \frac{\beta_x\epsilon_x}{\pi} - \frac{\beta_y\epsilon_y}{\pi}+
    (D_x\sigma_p)^2 
\right]
\label{th.8}
\end{equation}
where $\Delta_q$ is as defined above and we have set $D_y = 0$ because the vertical dispersion is small
in Booster.

We can write down a matrix equation using the above from the measurements
of $1$ to $n$ BPMs or $1$ to $n$ samples from $m$ BPMs
\begin{equation}
  \begin{pmatrix}
    \beta_x(1) & -\beta_y(1) & D_x^2(1)\\
    \beta_x(2) & -\beta_y(2) & D_x^2(2)\\
    \vdots & \vdots & \vdots\\
        \beta_x(j) & -\beta_y(j) & D_x^2(j)\\
    \vdots & \vdots & \vdots\\    
    \beta_x(n) & -\beta_y(n) & D_x^2(n)
\end{pmatrix}    
\begin{pmatrix}
  \epsilon_x/\pi\\
  \epsilon_y/\pi\\
  \sigma_p^2
\end{pmatrix}
=
\frac{a^2\phi_0}{2\sin\phi_0}
\begin{pmatrix}
  \Delta_q(1)\\
  \Delta_q(2)\\
  \vdots\\
  \Delta_q(j)\\
  \vdots\\  
  \Delta_q(n)
\end{pmatrix}
\label{th.9}
\end{equation}
where the $j$th row comes from the $j$th pickup or the $j$th sample.
If $n>3$ then we have a
non-square matrix on the lhs, which means that we have an
over-determined set of equations. This non-square matrix is easily
inverted using SVD methods. See for example, \textit{Mathematica}\/'s
\texttt{PseudoInverse[]\/} function. 

\subsection{Data Manipulations}

If we na\"ively perform the inversion in Eq.~(\ref{th.9}) to obtain the
emittances, we will find that we will get garbage. The keys for
getting sensible emittances are discussed below.

\begin{enumerate}
\item Theoretically, the gain on each BPM plate is required to be the
  same for Eq.~(\ref{th.9}) to work. Therefore, we have to correct the
  signal from the BPM plates to ensure that this condition is
  satisfied. Note: The choice of gain is critical for ensuring the
  emittance solutions are real and not complex. Details about our
  method is discussed in Ref.~\cite{cytan}.
\item Any DC offsets from each BPM plate when there is no beam.
\item The most important observation of Eq.~(\ref{th.9}) is that on the rhs, there is the difference
  $\beta_x\epsilon_x/\pi - \beta_y\epsilon_y/\pi$. This means that if
  we had a round beam and $\beta_x\approx\beta_y$ then this difference
  is close to zero. So, in this configuration, it may not be possible
  to actually extract out the emittances. The ideal situation for
  extracting out the emittances would be to have one of the $\beta$'s
  be a lot larger than the other, i.e.~either $\beta_x\gg\beta_y$ or
  $\beta_x\ll \beta_y$.
\end{enumerate}
For the Booster, the BPMs in the S locations are good for extracting
$\epsilon_x$ because $\beta_x(\SI{30}{m})\gg \beta_y(\SI{5}{m})$ while the BPMs at the L
locations are good for extracting $\epsilon_y$ because
$\beta_x(\SI{7}{m}) \ll \beta_y(\SI{20}{m})$.\footnote{We are using the symbols ``$\gg$'' and
  ``$\ll$'' loosely here.}  Therefore, Eq.~(\ref{th.9}) must contain
data from both L and S BPMs to get good $\epsilon_{x,y}$ solutions.
In this paper, we have paired the data from S01 and L01, S02 and L02,
etc.~for solving Eq.~(\ref{th.9}).

\subsection{Systematic Error Removal}

\begin{figure}[!htb]
   \centering
   \includegraphics*[width=\columnwidth]{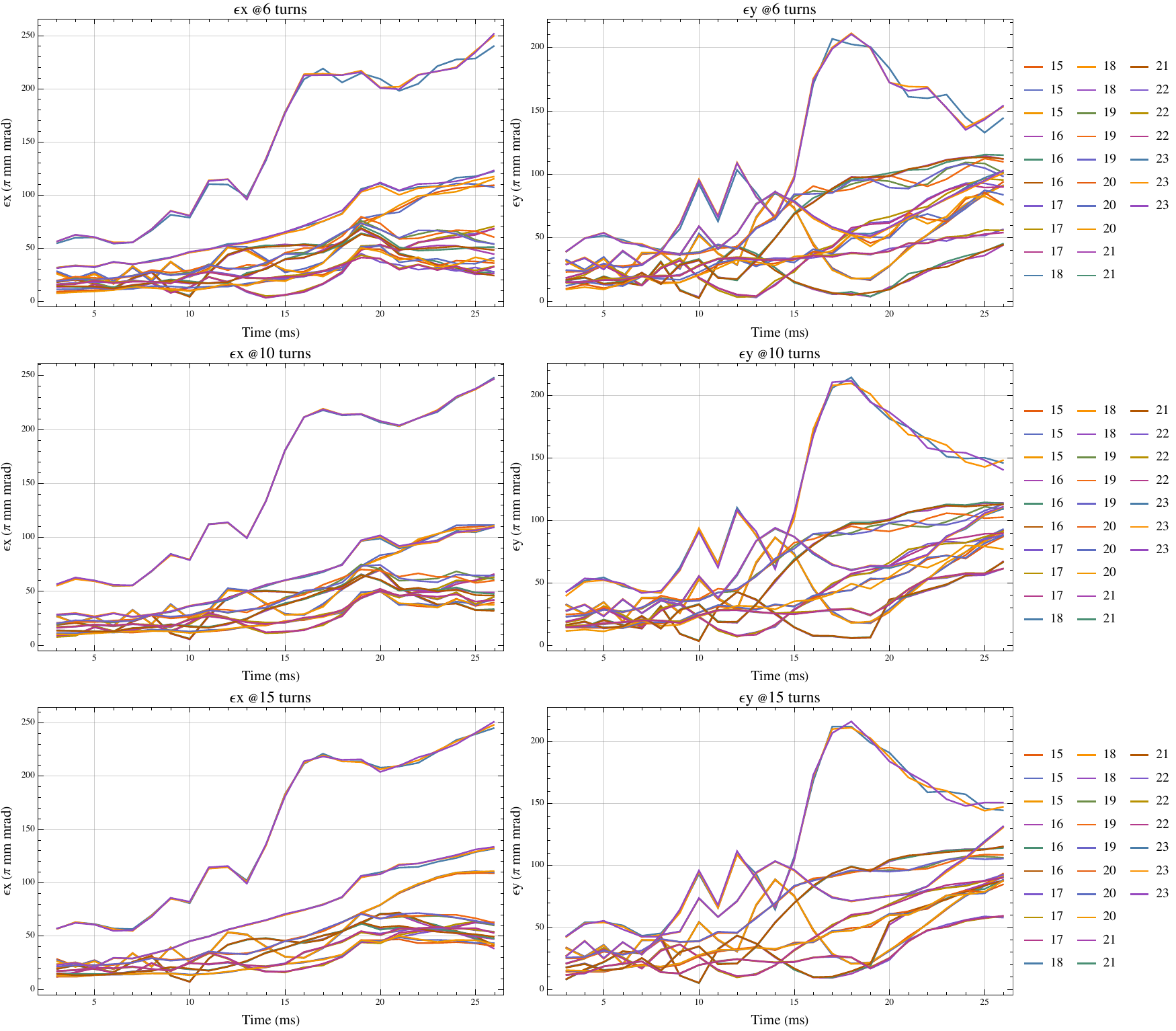}
   \caption{The systematic error of the emittances calculated with
     Eq.~(\ref{th.9}) for BPM pairs (L15, S15) to (L23, S23).}
   \label{compare_results_22Jun2023.pdf}
\end{figure}

We will find that even after applying the keys in the previous
section, the emittances found from Eq.~(\ref{th.9}) for every BPM pair
have different, but reproducible, systematic errors. The systematic
errors of each BPM pair (L15, S15) to (L23, S23) at three different
beam intensities at 6, 10 and 15 turns\footnote{The Linac beam is
  ``stacked'' in Booster in units of turns. Higher turns means higher
  intensity in Booster.} can be seen in
Fig.~\ref{compare_results_22Jun2023.pdf}. For each BPM pair, we have
collected data at each intensity thrice to show that the curves are
reproducible.

\begin{figure}[!htb]
   \centering
   \includegraphics*[width=\columnwidth]{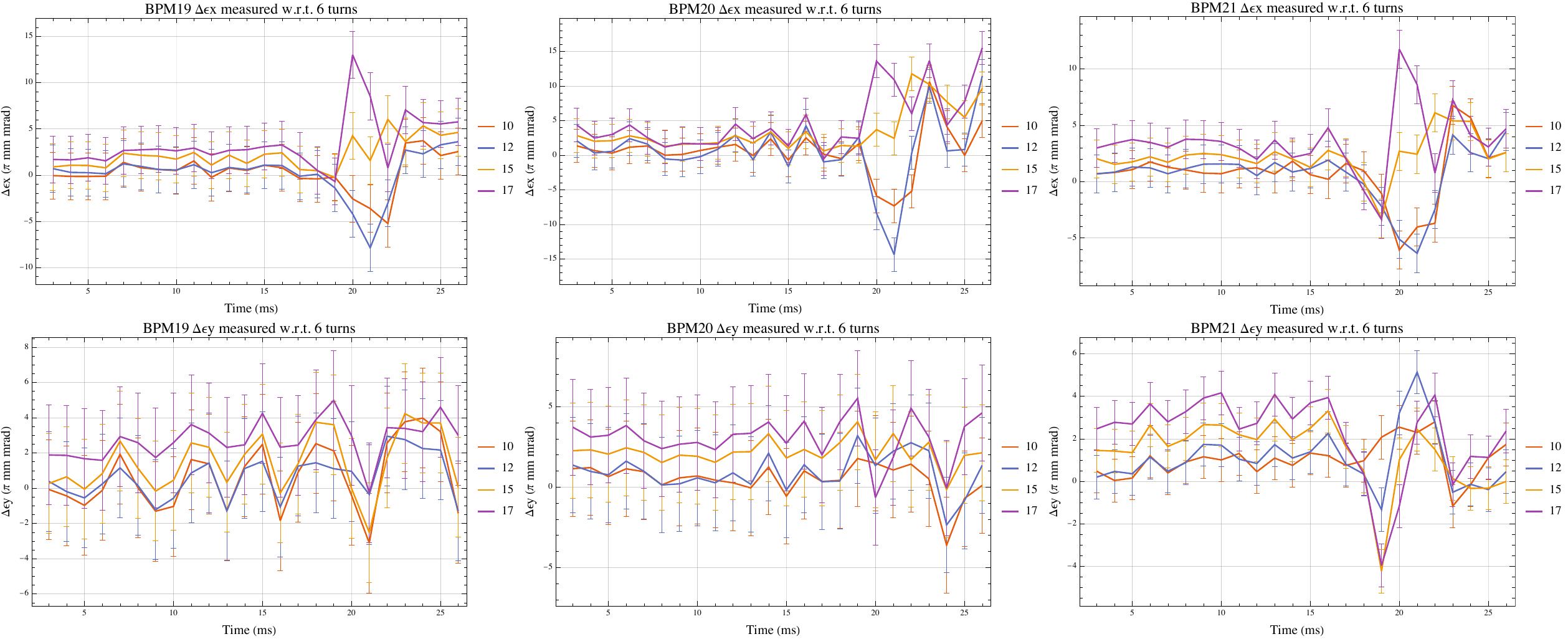}
   \caption{The horizontal and vertical emittance growth w.r.t.~6
     turns for BPM pairs (L19, S19) to (L21, S21) for 10, 12, 15 and
     17 turns are shown here.}
   \label{g19To21.pdf}
\end{figure}

Since the systematic error for each BPM pair is reproducible, we can
simply correct it by defining the 6 turn curve in
Fig.~\ref{compare_results_22Jun2023.pdf} to be the reference. We can
then take the difference between any $n$ turn curve with the 6 turn
curve to obtain the emittance growth w.r.t.~6 turns emittance. A
sample of the horizontal and vertical emittance growth w.r.t.~6 turns
for BPM pairs (L19, S19) to (L21, S21) for 10, 12, 15 and 17 turns are
shown in Fig.~\ref{g19To21.pdf}.

From this example, although the
curves from each BPM pair are not the same, we can identify some common features.
This leads us
to make the hypothesis that the differences are due to
underlying random noise that can be
averaged out \textit{if\/} our system is
ergodic. This means that we can average over the emittance from
all the BPM pairs (space average) which should be the same as
the averaged emittance at each BPM pair (time average).

\section{Comparison to IPM emittance}

The application of the ergodic hypothesis was the last step in the
analysis of our method.  At this point, we have to do some
verification. One way to do this is to compare the BPM and IPM
emittance measurements at different beam intensities.

\begin{figure}[!htb]
   \centering
   \includegraphics*[width=\columnwidth]{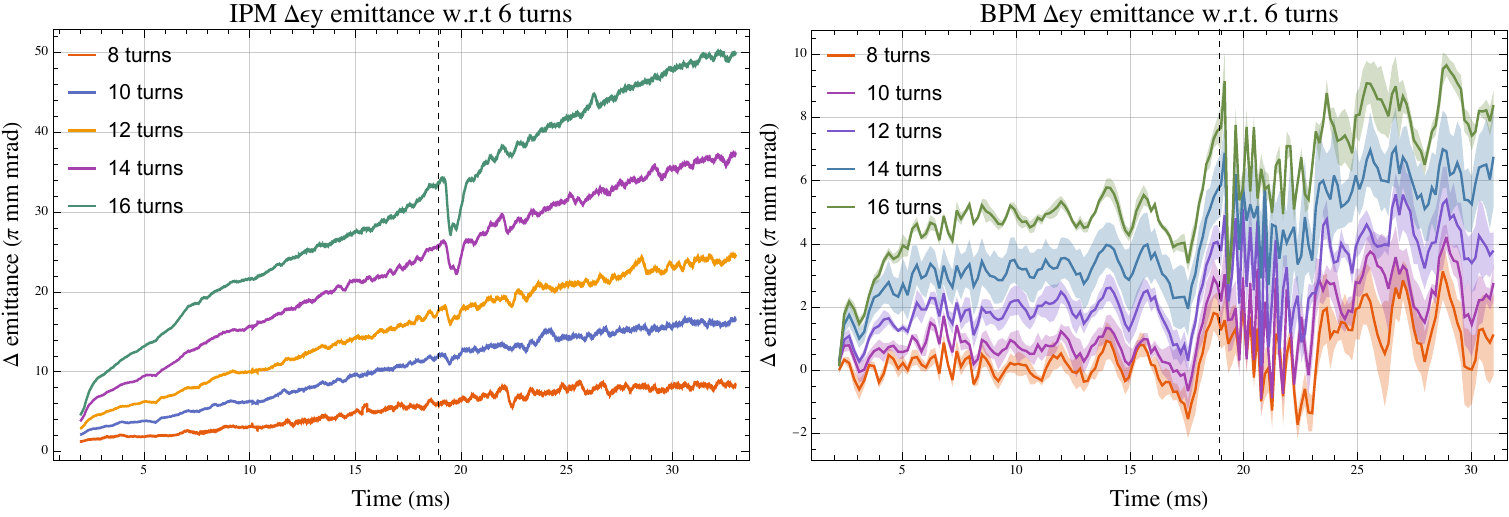}
   \caption{The vertical emittance growth w.r.t.~6 turns measured by
     the IPM (left) and the BPMs (right) for 8, 10, 12, 14, and 16
     turns. The dotted line is where the beam crosses
     transition. Note: the IPM measured emittance is too large; the shaded regions in the right plot are the
     statistical 1 sigma spread of the emittance at each intensity. Sectors 5, 14 and 22 were ignored because of bad BPM data.}
   \label{compare_ipm_bpm_y.pdf}
 \end{figure}

 For the first test, shown in Fig.~\ref{compare_ipm_bpm_y.pdf}, we
 will look at the vertical emittance at different intensities because
 the vertical IPM is the reliable channel.  Unfortunately, when we
 took the data, the IPM emittance results are at least a factor of 4
 -- 10 times too large and a linear growth which gets more steep as the
 intensity increases. We have not found any mechanism, including
 ecloud, that can cause this type of growth. On the other hand, the
 BPM emittance growth is a lot more realistic. The beam has an initial
 growth then flattens. A second growth happens after transition.

\begin{figure}[!htb]
   \centering
   \includegraphics*[width=\columnwidth]{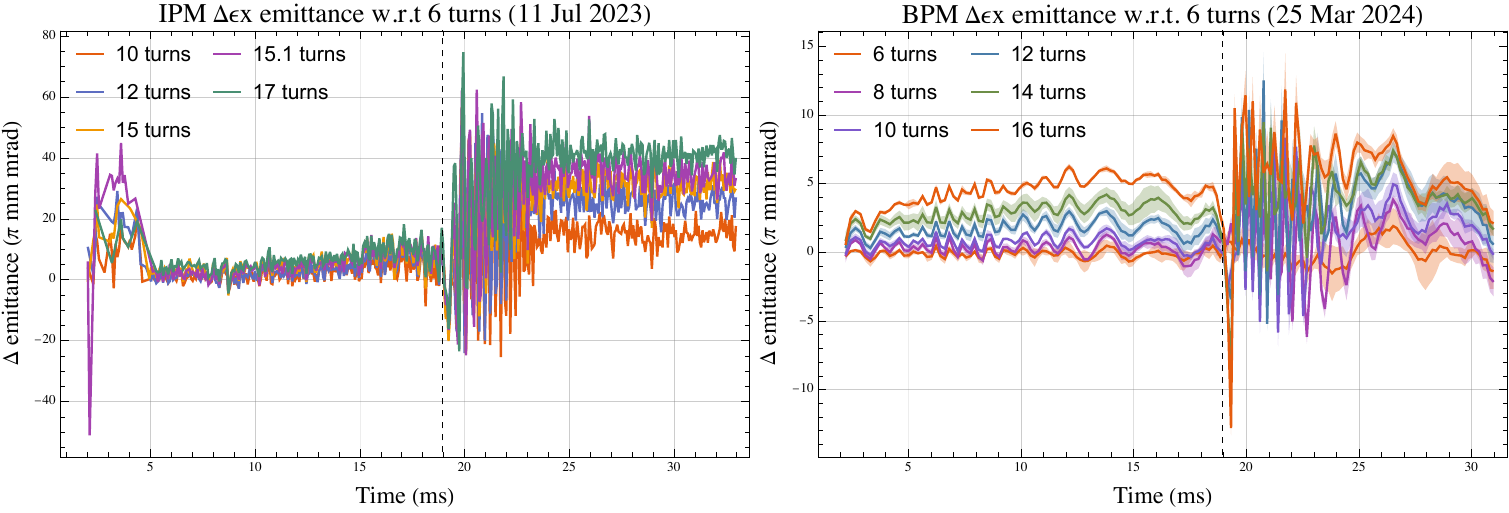}
   \caption{The horizontal emittance growth w.r.t.~6 turns measured by the IPM (left) and the
   BPMs (right) are shown here. The IPM emittance is about 4 times too
   large. Unfortunately, we can only compare
 data taken on different dates.}
   \label{compare_ipm_bpm_x.pdf}
 \end{figure}

For the second test, we will look at the horizontal emittance. On a
particular occasion (11 Jul 2023), the horizontal IPM functioned and
returned data that looked reasonable \textit{except\/} that it is too
big by at least a factor of 4. Due to the lack of horizontal IPM data,
we can only compare the IPM emittance growth taken on 11 Jul 2023 to the
the BPM emittance growth data taken on 25 Mar 2024 in
Fig.~\ref{compare_ipm_bpm_x.pdf}. Even though the data was taken about
7 months apart, they look similar, with the emittance oscillating
after transition. 

\section{CONCLUSION}

We have improved the extraction of emittance from the BPM quadrupole
mode by using multiple BPMs and with better data analysis. The
compromise is that we do not measure absolute emittance but the
emittance growth w.r.t.~some reference. We have found that the
emittance growth extracted from the BPM quadrupole mode looks more
sensible than those measured by the IPMs. As part of making this
method easy to use and without any expert tuning, we have written a
python GUI to interact with the user and a \texttt{C++} program
backend that can calculate Booster's emittance from real
time BPM data. If this method proves to be as robust as we believe it to be, it
opens up the ability of measuring emittances in nearly all rings that
have 4 plate BPMs.

\section{ACKNOWLEDGMENTS}

We would like to thank S. Chaurize, K. Triplett and J. Kuharik of the
PS/Booster group who helped us take
data. J. Diamond of the Controls group,  R. Santucci and D. Steincamp
of the Instrumentation group who helped us access the BPM
frontends. V. Kapin of PS/Physics and R. Thurman-Keup of
Instrumentation who helped us get the raw IPM data.

%
% only for "biblatex"
%
\ifboolexpr{bool{jacowbiblatex}}%
	{\printbibliography}%
	{%
	% "biblatex" is not used, go the "manual" way
	
	%\begin{thebibliography}{99}   % Use for  10-99  references
	
} % end \ifboolexpr

%
% for use as JACoW template the inclusion of the ANNEX parts have been commented out
% to generate the complete documentation please remove the "%" of the next two commands
% 
%%%%\newpage

%%%%\include{annexes-Letter}

\end{document}